\documentclass{llncs}   

\usepackage{amssymb}
\usepackage{amsmath}

\usepackage{makeidx}  
\usepackage{listings}
\usepackage{semantic}
\usepackage{wrapfig}
\usepackage{color}
\usepackage{url}
\DeclareMathAlphabet{\mathsc}{OT1}{cmr}{m}{sc}

\newcommand{\fname}[1]{\text{#1}} 
\newcommand{\rname}[1]{\mathrel{\fname{#1}}}
\newcommand{\oname}[1]{\mathit{#1}} 
\newcommand{\subject}{\Phi}
\newcommand{\ER}{\fname{ER}}

\newcommand{\ERmeet}{\mathrel{\wedge}}

\newcommand{\PT}{\fname{PT}}
\newcommand{\class}[1]{{[}#1{]}}
\newcommand{\erpart}[1]{
                        {\mathcal E}(#1)}
\newcommand{\know}[2]{\fname{K}_{#1}(#2)}
\newcommand{\kspace}[2]{{\mathcal K}_{#1}(#2)}

\newcommand{\sat}[1][{}]{\vdash_{#1}}
\newcommand{\psat}[2]{\vdash^{#2}_{#1}}
\newcommand{\korder}[1][{}]{\preceq_{\fname{#1}}}
\newcommand{\eptype}[2]{{\mbox{Type-}#1}({#2})}
\newcommand{\tuple}[1]{\langle #1 \rangle}
\newcommand{\Sys}[1]{\fname{Sys}(#1)}
\newcommand{\erased}{\eta}
\newcommand{\All}{\fname{All}}
\newcommand{\Id}{\fname{Id}}

\lstdefinelanguage{WHILE}{
  keywords={output, mod, div, to, int, while, if, then, else, for, on, do, skip},
}

\begin{document}
%
%

\title{A Semantic Hierarchy for Erasure Policies}
\titlerunning{Erasure Policies}  
%
\urldef{\mailsa}\path|{tedesco,dave}@chalmers.se|  
\urldef{\mailsb}\path|seb@city|

\author{Filippo Del Tedesco\inst{1} \and Sebastian Hunt\inst{2} \and David Sands\inst{1}}

\institute{Chalmers University of Technology, Sweden
\and 
  City University London 
}
\maketitle 

\authorrunning{Del Tedesco et al.} 
%
%
\begin{abstract}
We consider the problem
of logical data erasure,
contrasting with physical erasure
in the same way that end-to-end information flow control contrasts with access control.
We present a semantic hierarchy for erasure policies,
using a possibilistic knowledge-based semantics to define policy satisfaction such
that there is an intuitively clear upper bound on what information
an erasure policy permits to be retained. Our hierarchy allows
a rich class of erasure policies to be expressed, taking account of
the power of the attacker,
how much information may be retained, and under what conditions it may be
retained.
While our main aim is to specify erasure policies, the semantic framework
allows quite general information-flow policies to be formulated for a variety
of semantic notions of secrecy.
\end{abstract}


\section{Introduction}


Erasing data can be difficult for many reasons. As an example, recent
research on SSD-drives has shown that the low-level routines for
erasing data often inadvertently leave data behind \cite{Wei+:Reliably}.  This is
due to the fact that information on an SSD (in contrast to a more
conventional magnetic hard drive) gets copied to various parts of memory
in order to even out wear. 
The naive firmware sanitisation routines 
 do not have access to the movement-history of data, and so leave potentially large amounts of data behind.  

 This paper is not focused on low-level erasure \emph{per se}.  The
 requirement that data is used but not retained is commonplace in many
 non hardware-specific scenarios. As an everyday example consider the
 credit card details provided by a user to a payment system. The
 expectation is that card details will be used to authorise payment,
 but will not be retained by the system once the transaction is
 complete.

An erasure policy describes such a limited use of a piece of
data. 
But what does it mean for a system to correctly erase some piece of
data?  One natural approach taken here is to view erasure
as an information-flow concept -- following \cite{Chong:Myers:CSFW05}. To erase something means that after the point of erasure there is no information flowing from the original data to observers of the system. 
This gives
a natural generalisation of the low-level concept of \emph{physical
  erasure} to what one might call \emph{logical erasure}. Logical
erasure specifies that a system behaves \emph{as if} it
has physically erased some data \emph{from the viewpoint of a
  particular observer}. The observer viewpoint is more than just a way
to model erasure in a multi-level security context (as in \cite{Chong:Myers:CSFW05}).  To understand the
importance of the attacker viewpoint, consider a system which receives
some data subject to an erasure policy. The system then
receives a random key from a one-time pad and XORs it with the
secret. The key is then overwritten with a constant.  
Does such a system erase the data?
The answer, from an information-flow perspective, depends on what the
observer (a.k.a. \emph{the attacker}) can see/remember about the execution. 
An
attacker who can see the exact final state of the system (including the
encrypted data) and nothing more, cannot deduce
anything about the subject data, and so we can conclude that it is
erased for that attacker. But if the attacker could also observe the
key that was provided, then the system is not erasing. Different situations may need to model different attacker powers. 

In practice the concept of erasure is a subtle one in which many
dimensions play a role. This is analogous to the various ``dimensions'' of declassification \cite{Sabelfeld:Sands:Dimensions}.
 In this paper we develop a
semantic model for erasure which can account for
different \emph{amounts} of erasure, covering the situation where some
but not necessarily all information about the subject is removed, and
different varieties of \emph{conditional erasure}, which describes
both what is erased, and under what conditions.

The contribution of this work is to identify
(Section~\ref{sec:examples}) and formalise
(Section~\ref{sec:hierarchy}) a hierarchy of increasingly expressive
erasure policies which captures various dimensions of erasure.  To do
this we build on a new possibilistic information-flow model
(Section~\ref{sec:abstractmodel}) which is parameterised by (i) the
\emph{subject} of the information flow policy (e.g. the data to be
erased), (ii) the attacker's observational power. This is done taking
into account the \emph{facts} that an attacker might be interested to
learn, and the \emph{queries} which he can or will be able to answer
about the subject.

This is a pre-print, extended version of the work which includes proofs. The final publication is available at www.springerlink.com. 
\section{Erasure Case Studies}

\label{sec:examples}

We consider a series of examples of erasing systems which differ according to
the way they answer the following questions:
\begin{enumerate}
\item \emph{How much} of the erasure subject is erased?
\item \emph{Under which conditions} is erasure performed? 
\end{enumerate}
The examples are presented via simple imperative psudocode.  We
emphasise that the examples themselves are not intended to be
realistic programs -- they serve to motivate simply and intuitively 
various types of erasure policy that we will formalise in a more
abstract setting in Section~\ref{sec:hierarchy}.  

\subsection{Total erasure}
Consider a ticket vending machine using credit cards as the payment method.
A partial implementation, in simplified form, is shown in Listing~\ref{listing:total}.
\begin{wrapfigure}{l}{0.48\textwidth} 
\begin{lstlisting}[
caption={Ticket vending machine, total and unconditional erasure},
label={listing:total}
]
get(cc_number);
charge(ticket_cost,cc_number);
log(current_time());
cc_number=null;
\end{lstlisting}
\end{wrapfigure}
Line~1 inputs the card number;
line~2 executes the payment transaction;
line~3 writes the transaction time to a log for audit purposes;
line~4 deletes the card number.

This is an example of an erasing program: once line~4 is executed,
the card number has been erased from the system.
This statement can be refined further with respect to our original questions:
1) the system is \emph{totally} erasing
(no information about the card number is retained) and
2) erasure occurs \emph{unconditionally}, since control flow always reaches line 4.

\subsection{Partial erasure}
Consider a variant of the vending machine (Listing~\ref{listing:partial})
which logs the last four digits of the card number of each transaction,
enabling future confirmation of transactions in response to user queries.
The difference to Listing~\ref{listing:total} is in line~3, where additional data is written to the log.

\begin{wrapfigure}{l}{0.6\textwidth} 
\begin{lstlisting}[
caption={Ticket vending machine, partial and unconditional erasure},
label={listing:partial}]
get(cc_number);
charge(ticket_cost,cc_number);
log(current_time(),last4(cc_number));
cc_number=null;
\end{lstlisting}
\end{wrapfigure}
With this change, line~4 no longer results in total erasure since, even after
\lstinline!cc_number!
is overwritten, the last four digits of the card number are retained in the log. \newline

\subsection{Low dependent erasure}
Consider a further elaboration of the vending machine example
(Listing~\ref{listing:low-dependent})
which allows the user to \emph{choose} whether the
last four digits are retained.
\begin{wrapfigure}{r}{0.7\textwidth}
\begin{lstlisting}[caption={Ticket vending machine, low dependent erasure},label={listing:low-dependent},numbers=right]
get(cc_number);
charge(ticket_cost,cc_number);
get(choice);
if choice="Allow"
 then log(current_time(),last4(cc_number));
 else log(current_time());
cc_number=null;
\end{lstlisting}
\end{wrapfigure}
In line~3 the program acquires the user choice,
then it either proceeds as Listing~\ref{listing:partial}
or as Listing~\ref{listing:total}, according to the choice.
Now the question about how much information is erased has two different answers,
depending on the second user input.
Since this dependency is not related to the erasure subject itself,
we call this \emph{low dependent} erasure. 


\subsection{High dependent erasure}

Suppose there is a brand of credit cards, \fname{StealthCard}, 
which only allows terminals enforcing a strict confidentiality policy
to be connected to their network.
\begin{wrapfigure}{r}{0.53\textwidth}
\begin{lstlisting}[caption={Ticket vending machine, high dependent
erasure},label={listing:high-dependent},
 numbers=right]
get(cc_number);
charge(ticket_cost,cc_number);
if (cc_number is in StealthCard)
  then log(current_time());
  else get(choice);
    if choice="Allow"
      then log(current_time(),
               last4(cc_number));
      else log(current_time())
cc_number=null;
\end{lstlisting}
\end{wrapfigure}
This requires a further refinement of the program
(Listing~\ref{listing:high-dependent}),
since \fname{StealthCard} users are not permitted a choice for the logging option.   
At line~3 the credit card number is inspected and, if it is a \fname{StealthCard},
the system proceeds like \ref{listing:total}.

Compared to the previous case, this example has an additional layer of dependency,
since the amount of data to be erased is itself dependent on the erasure subject.
We refer to this as \emph{high dependent} erasure.

%
%

%
%
%
%

\section{An abstract model of information flow}
\label{sec:abstractmodel}

We formalise erasure policies as a particular class of information flow policies. In this section we define the basic building blocks for describing such policies. 
We consider trace-based (possibilistic) models of system behaviour and we interpret
information flow policies over these models.
We make the standard conservative assumption that the attacker has perfect knowledge of the system model.

Our definitions are based directly on what an attacker can deduce about
an erasure subject from observations of system behaviour. In this respect our model
is close in spirit
to Sutherland's multi-level security property of nondeducibilty \cite{Sutherland:NCSC86}.
However, we are not directly concerned with multi-level security
and, in a number of ways, our model is more abstract than non-deducibility.
For example, McLean's criticism \cite{McLean:Secure:Models:Inflow}
of nondeducibility (that it fails
to incorporate an appropriate notion of causality)
does not apply, since our notion of the ``subject'' of a policy
is general enough to incorporate temporal dependency if required.
On the other hand, our model \emph{is} open to the criticism of nondeducibilty
made by Wittbold and Johnson \cite{Wittbold:Johnson:Information}
with regard to interactive environment behaviours.
Adapting our work using the approach of
\cite{Wittbold:Johnson:Information}
(explicit modelling of user strategies) remains a subject for future work.
A more radical departure from the current work, though still possibilistic,
would be to take
a process-calculus approach \cite{Focardi:Gorrieri:Classification}.

\subsection{Trace models}

The behavioural ``atom'' in our framework is the \emph{event}
(in our examples this will typically be an input ($?v$) or output ($!v$)
but internal computation steps can be modelled in the same way).
Traces, ranged over by $s,t,s_1,t_1$, etc,
are finite or countably infinite sequences of events.
We write $t.e$ for the trace $t$ extended with event $e$ and we write
$s.t$ for the concatenation of traces $s$ and $t$.
%
In what follows we assume given some set $T$ of traces.

A \emph{system} is considered to be a set $S \subseteq T$
(the assumption is that $S$ is the set of maximal traces of the system being modeled).
Certain parts
of system behaviour will be identified as the \emph{subject} of our policies and we define
these parts by a function $\subject : T -> D$, for some set $D$
(typically, $\subject$ will be a projection on traces). For a confidentiality property the subject might represent the secret that we are trying to protect (an input or a behaviour of a classified agent).
For erasure the subject will be the input which is to be erased.

Given a system $S$, we denote by $\subject(S)$ the subset of $D$ relevant for $S$:
\[
	\subject(S) = \{ \subject(t) | t \in S \}
\]
We call this the \emph{subject domain} of $S$. 
Let $\Sys{V}$ be the set of all systems with subject domain $V$.
Our flow policies will be specific to systems with a given subject domain.

\subsection{Equivalence relations and partitions}
The essential component of a flow policy is a \emph{visibility} policy
which specifies how much an attacker should be allowed to learn about the
subject 
of a system by observing its behaviour.
Following a standard approach in the information flow literature
-- see, for example \cite{Landauer:Redmond:CSFW93,Sabelfeld:Sands:Per} --
we use equivalence relations for this purpose.
A flow policy for systems in $\Sys{V}$ is $R \in \ER(V)$,
where $\ER(V)$ denotes the set of all equivalence relations on $V$.
The intention is that attackers 
should not be able to distinguish between
subjects which are equivalent according to $R$.
An example is the ``have the same last four digits'' relation,
specifying that the most an attacker should be allowed to learn
is the last four digits of the credit card number (put another way,
all cards with the same last four digits should look the same to the attacker).
%

In what follows we make extensive use of two key, well known facts about
equivalence relations:
\begin{itemize}

\item
The set of equivalence relations on $V$, ordered
by inclusion of their defining sets of pairs, forms a complete lattice, with
the identity relation (which we denote $\Id_V$) as the bottom element,
and the total relation (which we denote $\All_V$) as the top.

\item
The set of equivalence relations on $V$ is
in one-one correspondence with the set of \emph{partitions} of $V$, where each
of the disjoint subsets making up a partition
is an equivalence class of the corresponding equivalence relation.
We write $\PT(V)$ for the set of all partitions of $V$.
Given $P \in \PT(V)$, we write ${\erpart{P}}$ for the corresponding
equivalence relation:
$v_1 \erpart{P} v_2 \mbox{ iff } \exists X \in P.v_1, v_2 \in X$.
In the other direction, given
$R \in \ER(V)$ and $v \in V$ we write $\class{v}_R$ for the $R$-equivalence class
of $v$:
\(
	\class{v}_R = \{ v' \in V | v' \mathrel{R} v \}
\).
We write $\class{R}$ for the partition corresponding to $R$:
\(
	\class{R} = \{ \class{v}_R | v \in V \}
\).

\end{itemize}
In the current context, the significance of $R_1 \subseteq R_2$ is that
$R_1$ is more discriminating - i.e., has smaller equivalence classes - than
$R_2$. Hence, as visibility policies, $R_1$ is more permissive than $R_2$.
The lattice operation of interest on $\ER(V)$ is \emph{meet},
which is given by set intersection. Given a family of equivalence relations
$\{ R_i \}_{i \in I}$, we write their meet as $\bigwedge_{i \in I}R_i$
(the least permissive equivalence relation which is nonetheless
more permissive than each $R_i$).

The order relation on partitions corresponding to subset inclusion
on equivalence relations will be written $\korder[ER]$, thus
$\class{R_1} \korder[ER] \class{R_2}$ iff $R_1 \subseteq R_2$.
We overload the above notation for meets of partitions in this isomorphic lattice:
\(
	\bigwedge_{i \in I}P_i = \class{\bigwedge_{i \in I}{\erpart{P_i}}}
\).

\subsection{Attacker models and K-spaces}
As discussed in the introduction,
whether or not a system satisfies a policy will depend on what is observable
to the attacker. We specify an \emph{attacker model}
as an equivalence relation on traces, $A \in \ER(T)$.
Note that this
is a passive notion of attacker - attackers can observe but not interact with
the system.

To compare what the attacker actually learns about the subject with what the visibility
policy permits, we define, for each attacker observation $O \in \class{A}$, the corresponding
\emph{knowledge set} $\know{S}{O} \subseteq V$,
which is the set of possible subject 
values which the attacker can deduce from making a given observation\footnote{A more reasonable but less conventional terminology would be to call this an \emph{uncertainty set}.}:
\(
	\know{S}{O} = \{ \subject(t) | t \in O \cap S \}
\).
   
The \emph{K-space of $A$ for $S$}, denoted $\kspace{S}{A}$, is the collection of all the
attacker's possible (ie non-empty) knowledge sets when observing $S$:
\[
	\kspace{S}{A} = \{ \know{S}{O} | O \in \class{A}, O \cap S \neq \emptyset \}
\]

\begin{lemma}
Let $S \in \Sys{V}$ and $A \in \ER(V)$. Then the K-space of $A$ for $S$ \emph{covers} $V$,
by which we mean that every member of $\kspace{S}{A}$ is non-empty and
$\bigcup\kspace{S}{A} = V$.
\end{lemma}
From now on, for a given $V$, we use the term K-space to mean any collection of sets which covers $V$.

In the special case that a system's behaviour is a \emph{function} of the
subject, each K-space will actually define an equivalence relation on $V$:
\begin{proposition}
Say that $S \in \Sys{V}$ is \emph{functional} just when, for all $t, t' \in S$,
$t \neq t' => \subject(t) \neq \subject(t')$.
In this case, for all $A \in \ER(T)$,  $\kspace{S}{A}$ partitions $V$.
\end{proposition}


When $S$ is functional, the K-space $\kspace{S}{A}$, being a partition, can be interpreted
as the equivalence relation ${\erpart{\kspace{S}{A}}}$.
So, in the functional case there is a straightforward way to compare a visibility
policy with an attacker's K-space: we say that the policy $R$ is satisfied just when
$R$ is more discriminating than this induced equivalence relation.
Formally, when $S$ is functional,
\emph{$S$ satisfies $R$ for attacker $A$}, written $S \sat[A] R$, just when
$R \subseteq {\erpart{\kspace{S}{A}}}$ or, equivalently:
\begin{equation}\label{eqn:functional-sat}
	S \sat[A] R \mbox{ iff } \class{R} \korder[ER] \kspace{S}{A}
\end{equation}
We now consider how to extend this definition to the general case, in which
a system has other inputs apart from the policy subject.

\subsection{Comparing K-Spaces: facts and queries}
In general, a system's behaviour may depend on events which are neither part
of the policy subject nor visible to the attacker. In this case,
the attacker's knowledge of the subject need not be deterministic, resulting
in a K-space which is not a partition.
This raises the question: when is one K-space ``more discriminating'' than another?  

Here we motivate a variety of orderings by considering some basic
modes in which an attacker can use observations to make deductions
about the subject of a system:
\begin{description}
\item[Facts]  A fact $F$ is just a set of values. A given knowledge set
  $X$ \emph{confirms fact $F$} just when $X \subseteq
  F$. Dually, $X$ \emph{has uncertainty} $F$ when $F \subseteq X$. For example a fact of interest (to an attacker) might be the set of
``Platinum'' card numbers. In this case an observation might confirm to the attacker that
a card is a Platinum card by also revealing exactly which platinum
card it is. 
For a given K-space $K$ we then say that
\begin{itemize}
\item $K$ \emph{can confirm} $F$ if there exists some $X \in K$ such that $X$ confirms $F$.
\item $K$ \emph{can have uncertainty} $F$ if there exists some $X \in K$ such that $X$ has uncertainty $F$.
\end{itemize}
\item[Queries] A query $Q$ is also just a set of values. We say that 
a given knowledge set $X$ \emph{answers query
  $Q$} just when either $X \subseteq Q$ or $X \subseteq V\setminus Q$. For a given K-space $K$ we then say that
\begin{itemize}
\item $K$ \emph{will answer} $Q$ if for all $X \in K$, $X$ answers $Q$, and 
\item $K$ \emph{can answer} $Q$ if there exists some $X \in K$ such that
$X$ answers $Q$. 
\end{itemize}
\end{description}
In a possibilistic setting it is natural to focus on those ``secrets''
which it is \emph{impossible} for a given system to reveal, where
revealing a secret could mean either confirming a fact or answering a query.
Two of the four K-space properties defined above
have an immediate significance for this notion of secrecy:
\begin{itemize}
\item Say that $S$
\emph{keeps fact $F$ secret from attacker $A$} iff there are no runs
of $S$ for which $A$'s observation confirms $F$,
i.e., iff:
\( \neg(\kspace{S}{A} \rname{can confirm} F) \).

\item Say that $S$
\emph{keeps query $Q$ secret from attacker $A$} iff there are no runs
of $S$ for which $A$'s observation answers $Q$,
i.e., iff:
\( \neg(\kspace{S}{A} \rname{can answer} Q) \).

\end{itemize}
The possibilistic secrecy significance of ``has uncertainty'' and ``will answer''
is not so clear. However, as we will show, we are able to define flow policies
and a parameterized notion of policy satisfaction
which behaves well with respect to all four properties.

Using the ability of a K-space to confirm facts and answer queries,
we can order systems in different ways, where a ``smaller''
K-space (ie one lower down in the ordering) allows the attacker to make
more deductions (and so the system may be regarded as less secure).
%
Define the following orderings between K-spaces:
\begin{description}
\item[Upper:]
$K_1 \korder[U] K_2 \mbox{ iff } \forall F . K_2 \rname{can confirm} F => K_1 \rname{can confirm} F$.
Note that $K_1 \korder[U] K_2$ iff $K_2$ keeps more facts secret than $K_1$.
\item[Lower:] 
$K_1 \korder[L] K_2 \mbox{ iff } \forall F . K_1 \rname{can have uncertainty} F => K_2 \rname{can have uncertainty} F$.
\item[Convex (Egli-Milner):] 
$K_1 \korder[EM] K_2 \mbox{ iff } K_1 \korder[U] K_2 \ERmeet K_1 \korder[L] K_2$.
\item[Can-Answer:]
$K_1 \korder[CA] K_2 \mbox{ iff } \forall Q 
. K_2 \rname{can answer} Q => K_1 \rname{can answer} Q$. %
Note that $K_1 \korder[CA] K_2$ iff $K_2$ keeps more queries secret than $K_1$.
\item[Will-Answer:]
$K_1 \korder[WA] K_2 \mbox{ iff } \forall Q
. K_2 \rname{will answer} Q => K_1 \rname{will answer} Q$.
\end{description}
It is straightforward to verify that these orders are reflexive and
transitive, but not anti-symmetric.
The choice of names for the upper and lower orders
is due to their correspondence with the powerdomain orderings
\cite{plotkin:powerdomain}: 
\begin{proposition}
\begin{align*}
    K_1 \korder[U] K_2 & \mbox{ iff } \forall X_2 \in K_2. \exists X_1
    \in K_1 . X_1 \subseteq X_2
\\
K_1 \korder[L] K_2 & \mbox{ iff }
    \forall X_1 \in K_1. \exists X_2 \in K_2 . X_1 \subseteq
    X_2
  \end{align*}
\end{proposition}

We can compare the K-space orders
1) unconditionally,
2) as in the case of policy satisfaction, when we are comparing a partition with a K-space,
and,
3) when the K-spaces are both partitions,
yielding the following results:
\begin{proposition}\label{prop:korders}
  \begin{enumerate} 


	\item ${\korder[EM]} \subsetneq {\korder[L]} \subsetneq {\korder[WA]}$ and
			${\korder[EM]} \subsetneq {\korder[U]} \subsetneq {\korder[CA]}$.


\item Additionally, when $P$ is a partition:
	$P \korder[CA] K => P \korder[WA] K$ (the reverse implication does not hold in general).


\item $\korder[ER]$, $\korder[EM]$, $\korder[L]$, and $\korder[WA]$
all coincide on partitions.
Furthermore, when $P_1$ and $P_2$ are partitions:
$P_1 \korder[ER] P_2 => P_1 \korder[U] P_2 => P_1 \korder[CA] P_2$ (the reverse implications do not hold in general).

\end{enumerate}
\end{proposition}
These orderings give us a variety of ways to extend the definition
of policy satisfaction from functional systems (Equation~\ref{eqn:functional-sat})
to the general case. The choice will depend on the type of security
condition (eg protection of facts versus protection of queries) which we wish
to impose.






\section{The policy hierarchy}
\label{sec:hierarchy}

We specify a three-level hierarchy of erasure policy types.
All three types of policy use a structured collection of equivalence relations on
the subject domain to define what information should be erased.
A key design principle is that, whenever a policy permits part of the erasure subject to
be retained, this should be \emph{explicit},
by which we mean that it should be captured
by the conjunction of the component equivalence relations.

For each type of policy, we define a satisfaction relation, parameterized by a
choice of K-space ordering
$o \in \{ \oname{U}, \oname{L}, \oname{EM}, \oname{CA}, \oname{WA} \}$.

Assume a fixed policy subject function $\subject : T -> D$.
Given a subset $V \subseteq D$,
let $T_V = \{ t \in T | \subject(t) \in V \}$.
Note that if $S$ belongs to $\Sys{V}$ then $S \subseteq T_V$.

\subsection*{Type 0 policies} \label{sec:type0}
Type 0 policies allow
us to specify \emph{unconditional} erasure, corresponding to the two examples
shown in Section~\ref{sec:examples} in Listings \ref{listing:total} and \ref{listing:partial}.

A Type~0 erasure policy is just a visibility policy.
We write $\eptype{0}{V}$
for the set of all Type~0 policies for systems in $\Sys{V}$ (thus $\eptype{0}{V} = \ER(V)$).
The definition of satisfaction for a given attacker model $A$
and system $S$ uses a K-space ordering
(specified by parameter $o$)
to generalise the satisfaction relation of Equation~\ref{eqn:functional-sat}
to arbitrary (i.e., not-necessarily functional) systems:
\[
	S \psat{A}{o} R \mbox{ iff } \class{R} \korder[o] \kspace{S}{A}
\]
For functional systems note that, by Proposition~\ref{prop:korders},
choosing $o$ to be any one of
$\oname{EM}$, $\oname{L}$ or $\oname{WA}$
yields a notion of satisfaction
equivalent to Equation~\ref{eqn:functional-sat},
while $\oname{U}$ and $\oname{CA}$ yield strictly weaker notions.

\noindent\textit{Example}.
Consider the example in Listing~\ref{listing:partial}.
The subject domain is \fname{CC}, the set of all credit card numbers, and
(since the erasure subject is the initial input) the subject function is the first projection
on traces.
The policy we have in mind for this
system is that it should erase all but the last four digits of the credit card number. 
We extend this example so that it uses a method call
\texttt{erased()} to generate an explicit output event $\erased$
(signalling that erasure should have taken place)
followed by a dump of the program memory
(thus revealing all retained information to a sufficiently strong attacker).

\begin{wrapfigure}{l}{0.6\textwidth}
\begin{lstlisting}[
caption={Ticket vending machine, partial and unconditional erasure: extended},
label={listing:partial-extended}]
get(cc_number);
charge(ticket_cost,cc_number);
log(current_time(),last4(cc_number));
cc_number=null;
erased();
dump();
\end{lstlisting}
\end{wrapfigure}
If we restrict attention to systems (such as this one) where each run starts by
inputting a credit card number and eventually outputs
the erasure signal exactly once, we can assume a universe of traces $T$ such that
all $t \in T$ have the form $t = ?\mathit{cc}.s.\erased.s'$,
where $s, s'$ are sequences 
not including $\erased$.
Let $S$ be the trace model for the above system.
The required visibility policy is the equivalence relation $\fname{L4} \in \ER(\fname{CC})$ which equates
any two credit card numbers sharing the same last four digits.
An appropriate attacker model is the attacker
who sees nothing before the erasure event and everything afterwards.
Call this the \emph{simple erasure attacker}, denoted $\fname{AS}$:
\[
	\fname{AS} = \{ (t_1, t_2) \in T \times T |
		\exists s_1, s_2, s_3.\  t_1 = s_1.\erased.s_3 \  \ERmeet \ t_2 = s_2.\erased.s_3 \}
\]
Informally, it should be clear that, for each run of the system,
$\fname{AS}$ will learn the last four digits of the credit card which was input,
together with some other log data (the transaction time) which is independent of the
card number. Thus the knowledge set on a run,
for example, where the card number ends 7016, would be the set of all card numbers
ending 7016. The K-space in this example will actually be exactly the partition
$\class{\fname{L4}}$,
hence $S$ does indeed satisfy the specified policy:
$S \psat{\fname{AS}}{o} \fname{L4}$
for all choices of $o$.
From now on, we write just $S \sat[A] R$ to mean that it holds
for all choices of ordering
(or, equivalently, we can consider $\sat[A]$ to be shorthand for
$\psat{A}{EM}$, since $\oname{EM}$ is the strongest ordering).

\subsection*{Type 1 policies}
Type~1 policies allow us to specify
``low dependent'' erasure (Section~\ref{sec:examples}, Listing~\ref{listing:low-dependent}),
where different amounts may be erased on different runs, but where the erasure condition
is independent of the erasure subject itself.

For systems in $\Sys{V}$ the erasure condition is specified
as a partition $P \in \PT(T_V)$. 
This is paired with a function $f : P -> \eptype{0}{V}$, which associates
a Type~0 policy with each element of the partition.
Since the domain of $f$ is determined by the choice of $P$, we use a dependent type
notation to specify the set of all Type~1 policies:
\[
	\eptype{1}{V} = \tuple{P:\PT(T_V), P -> \ER(V)}
\]
Because we want to allow only low dependency
-- i.e., the erasure condition must be independent of the erasure subject --
we require that $P$ is \emph{total for $V$}, by which we mean:
\[
	\forall X \in P . \subject(X) = V
\]
This means that knowing the value of the condition will not in itself rule out
any possible subject values. To define policy satisfaction
we use the components $X \in P$
to partition a system $S$ into disjoint sub-systems $S \cap X$ and check both that
each sub-system is defined over the whole subject domain $V$ (again, to ensure
low dependency) and that it satisfies the Type~0 policy for sub-domain $X$.
So, for a Type~1 policy
$\tuple{P,f} \in \eptype{1}{V}$,
an attacker model $A$,
and system $S \in \Sys{V}$, satisfaction is defined thus:
\[
	S \psat{A}{o} \tuple{P,f} \mbox{ iff } \forall X \in P . S_X \in \Sys{V} \ERmeet S_X \psat{A}{o} f\;X
\]
where $S_X = S \cap X$.

\noindent\textit{Example}.
  Consider the example of Listing~\ref{listing:low-dependent} extended
  with an erasure signal followed by a memory dump (as in our
  discussion of Type~0 policies above).  Let $S$ be the system model
  for the extended program.  We specify a conditional erasure policy
  where the condition depends solely on the user choice.  The erasure
  condition can be formalised as the partition $\fname{Ch} \in \PT(T)$
  with two parts, one for traces where the user answers ``Allow''
  (which we abbreviate to $a$) and one for traces where he doesn't: \(
  \fname{Ch} = \{ Y, \overline{Y} \} \),
  where $Y = \{ t \in T | \exists s,s_1,s_2 .\ t =
  s.?a.s_1.\erased.s_2 \}$ and $\overline{Y} = T \setminus Y$. For
  runs falling in the $Y$ component, the intended visibility policy is
  $\fname{L4}$, as in the Type~0 example above.  For all other runs,
  the intended policy is $\All_{\fname{CC}}$, specifying complete
  erasure. The Type~1 policy is thus $\tuple{\fname{Ch}, g}$ where $g
  : \fname{Ch} -> \ER(\fname{CC})$ is given by:
  \[
  g(X) = \left\{\begin{array}{lcl}
      \fname{L4} & \mbox{ if } X = Y \\
      \All & \mbox{ if } X = \overline{Y} \\
    \end{array}\right.
  \]
  Intersecting $Y$ and $\overline{Y}$, respectively, with the system
  model $S$ gives disjoint sub-systems $S_Y$ (all the runs in which
  the user enters ``Allow'' to permit retention of the last four
  digits) and $S_{\overline{Y}}$ (all the other runs).  Since the
  user's erasure choice is input independently of the card number, it
  is easy to see that both sub-systems are in $\Sys{\fname{CC}}$, that
  $S_Y \sat[\fname{AS}] \fname{L4}$, and $S_{\overline{Y}}
  \sat[\fname{AS}] \All$.  Thus $S \sat[\fname{AS}] \tuple{\fname{Ch},
    g}$.



The following theorem establishes that our ``explicitness'' design principle
is realised by Type~1 policies:
\begin{theorem}\label{theorem:explicitness-type1}
Let $\tuple{P,f} \in \eptype{1}{V}$ and $S \in \Sys{V}$ and $A \in \ER(T)$.
Let
$o \in \{ \oname{U}, \oname{L}, \oname{EM}, \oname{CA}, \oname{WA} \}$.
If $S \psat{A}{o} \tuple{P,f}$ then:
\[
	\class{\bigwedge_{X \in P}(f\;X)} \korder[o] \kspace{S}{A}
\]
\end{theorem}
\noindent\textit{Example}.
Consider instantiating the theorem to the policy $\tuple{\fname{Ch}, g}$ 
described above. Here the policy is built
from the two equivalence relations $\All$ and $\fname{L4}$; the
theorem tells us that the knowledge of the
attacker is bounded by the meet of these components (and hence
nothing that is not an explicit part of the policy) i.e., $\All \ERmeet
\fname{L4}$, which is equivalent to just $\fname{L4}$.

\subsection*{Type 2 policies}
Type 2 policies  are the most flexible policies we consider, allowing dependency on both
the erasure subject and other properties of a run.

Recall the motivating example from Section~\ref{sec:examples}
(Listing~\ref{listing:high-dependent}) in which credit card numbers in
a particular set (the StealthCards)
$\fname{SC} \subseteq \fname{CC}$ 
are always
erased, while the user is given some choice for other card numbers.
In this example, the dependency of the policy on the erasure subject
can be modelled by the partition 
$\fname{HC} = \{ \fname{SC}, \overline{\fname{SC}} \}$.
For each of these two cases,
we can specify sub-policies which apply only to card numbers
in the corresponding subsets. Since these sub-policies do not involve any further
dependence on the erasure subject, they can both be formulated as Type~1 policies
for their respective sub-domains.
In general then, we define the Type~2 policies as follows:
\[
	\eptype{2}{V} = \tuple{Q:\PT(V), W:Q -> \eptype{1}{W}}
\]

To define satisfaction for Type~2 policies, we use the components
$W \in Q$ to partition a system $S$ into sub-systems
(unlike the analogous situation with Type~1 policies,
we cannot intersect $S$ directly with $W$;
instead, we intersect with $T_W$).
To ensure that the \emph{only} dependency on the erasure subject is that described by $Q$,
we require that each sub-system $S \cap T_W$ is defined over the whole of the subject sub-domain $W$.
So, for a Type~2 policy
$\tuple{Q,g} \in \eptype{2}{V}$,
an attacker model $A$,
and system $S \in \Sys{V}$, satisfaction is defined thus:
\[
	S \psat{A}{o} \tuple{Q,g} \mbox{ iff } \forall W \in Q .
		S_W \in \Sys{W} \ERmeet S_W \psat{A}{o} g\;W
\]
where $S_W = S \cap T_W$.


To state the appropriate analogue of Theorem~\ref{theorem:explicitness-type1}
we need to form a conjunction of all the component parts of a Type~2 policy:
\begin{itemize}
\item In the worst case, the attacker will be able to observe which of the erasure
cases specified by $Q$ contains the subject,
hence we should conjoin the corresponding equivalence relation $\erpart{Q}$.
\item Each Type~1 sub-policy determines a worst case equivalence relation, as
defined in Theorem~\ref{theorem:explicitness-type1}. To conjoin these relations,
we must first extend each one from its sub-domain to the whole domain,
by appending a single additional equivalence class comprising all the ``missing'' elements:
given $W \subseteq V$ and $R \in \ER(W)$, define $R^{\dagger} \in \ER(V)$ by
$R^{\dagger} = R \cup \All_{V\setminus W}$.
\end{itemize}
\begin{theorem}\label{theorem:explicitness-type2}
Let $\tuple{Q,g} \in \eptype{2}{V}$ and $S \in \Sys{V}$ and $A \in \ER(T)$.
For any Type~1 policy $\tuple{P,f}$,
let $R_{\tuple{P,f}} = \bigwedge_{X \in P}(f\;X)$.
Let $o \in \{ \oname{U}, \oname{L}, \oname{EM}, \oname{CA}, \oname{WA} \}$.
If $S \psat{A}{o} \tuple{Q,g}$ then:
\[
	\class{\erpart{Q} \ERmeet \bigwedge_{W \in Q}R^{\dagger}_{(g\;W)}} \korder[o] \kspace{S}{A}
\]

%
\end{theorem}
\noindent\textit{Example}
We consider a Type~2 policy satisfied by
Listing~\ref{listing:high-dependent}, namely 
$\tuple{\fname{HC},h}$ where $\fname{HC}$ is the partition into Stealth and
non-Stealth cards (as above), and 
$h$ is defined as follows.

\newcommand{\SC}{\fname{SC}}
\newcommand{\nSC}{\overline{\SC}}

\[
\begin{array}[t]{lcl}
h(\SC) & = & \tuple{\{T_{\SC}\},\lambda x.\All_{\SC}}
\\
h(\nSC) & = &\tuple{\fname{Ch}, h_1}
\end{array}
\qquad
\begin{array}[t]{lcl}
h_1 (Y) & =   & \fname{L4}_{\nSC}
\\
h_1 (\overline{Y}) & =   & \All_{\nSC}
\end{array}
\]  
\noindent
The term $T_{\SC}$ denotes the set of traces which input a Stealth card number as first action. As in the example of Type~1 policy above, $Y$ is the set of (non-stealth) traces where the user gives
permission (``Yes'') to retain the last digits, $\overline{Y}$ is its  complement (relative to the set of non-stealth traces), and $\fname{Ch}$ is
the partition $\{Y,\overline Y \}$. The term $\fname{L4}_{\nSC}$ denotes the restriction of $L4$ to elements in $\nSC$. Instantiating Theorem~\ref{theorem:explicitness-type2} to this example
tells us that the attacker knowledge is bounded by 
\(
 {\erpart{\fname{HC}}} \ERmeet {\All_{\SC}^{\dagger}} \ERmeet
  \fname{L4}_{\nSC}^{\dagger} \ERmeet {\All_{\nSC}^{\dagger}},
\) which is just $\fname{L4}_{\nSC}^{\dagger}$.

\subsection{Varying the attacker model}
The hierarchy deals with erasure policies independently of any
particular attacker model. Here we make some brief remarks about
modelling attackers. Let us take the example of the erasure notion
studied in \cite{Hunt:Sands:ESOP08} where the systems are simple
imperative programs involving IO on public and secret channels. Then
the \emph{implicit} attacker model in that work is
unable to observe any IO events prior to the erasure point, and is able to observe
just the public inputs and outputs thereafter. (We note that \cite{Hunt:Sands:ESOP08}
also considers a policy
\emph{enforcement} mechanism which uses a stronger, state-based non-interference
property.)


Now consider the example of the one-time pad described in the introduction, codified in Listing~\ref{listing:key}.
Let system $S$ be the set of traces modelling the possible runs of the program
and let the subject be the first input in each trace.  
For the simple erasure attacker $AS$ (Section~\ref{sec:type0}),
unable to observe the key provided in line 2, the
\noindent
K-space will be $\{ V \} = \class{\All}$, hence
$S \sat[\fname{AS}] \All$. This is because the value of 
\lstinline!data! in the output does not inform the attacker about the initial value.
\begin{wrapfigure}{l}{0.34\textwidth}
\begin{lstlisting}[
caption={Key Erasure},
label={listing:key}]
get(data); 
get(key);
data := data XOR key;
key := null;
erased();
output(data);
\end{lstlisting}
\end{wrapfigure}
On the other hand, the attacker who can also observe the key
learns everything about the data from its encrypted value.\footnote{Note, however, that we cannot model the fact that certain functions are not (easily) invertible, so our attackers are always endowed with 
unbounded computational power.} 
So for this stronger attacker, using encryption to achieve erasure does not work,
and indeed policy satisfaction fails for this particular system.

If the attacker is strengthened even further, we arrive at a point where \emph{no}
system will be able to satisfy the policy. Intuitively, if an attacker can see the
erasure subject itself (or, more specifically, more of the erasure subject than
the policy permits to be retained) no system will be able to satisfy the policy.
In general, we say that a policy $p$ with subject domain $V$ (where $p$ may be
of any of Types 0,1,2) is \emph{weakly $o$-compatible} with attacker model $A$ iff
there exists $S \in \Sys{V}$ such that $S \psat{A}{o} p$
(we call this weak compatibility because it assumes that all $S \in \Sys{V}$
are of interest but in general there will be additional constraints
on the admissible systems). Clearly, to be helpful as
a sanity check on policies we need something a little more constructive than this.
For the special case of Type~0 policies and the upper ordering we have the following
characterisation:
\begin{lemma}
$R$ is weakly $\oname{U}$-compatible with $A$ iff
$\forall v \in V . \exists O \in \class{A} . \class{v}_R \subseteq \Phi(O)$.
\end{lemma}
Deriving analogues of this result (or at least sufficient conditions) of
more general applicability remains a subject for further work.

Finally, we note that, while our main aim has been to specify erasure policies,
by varying the attacker model appropriately, we can specify
quite general information-flow properties, not just erasure policies.
For example, by classifying events into High and Low and defining
the attacker who sees only Low events, we can specify non-interference properties.

%

\section{Related work}
We consider related work both directly concerned with erasure
and more generally with knowledge based approaches to information flow policies.

\paragraph{Erasure}
The information-flow perspective on erasure was introduced by Chong
and Myers \cite{Chong:Myers:CSFW05} and was studied in combination
with confidentiality and declassification. Their semantics is
based on an adaptation of two-run noninterference definitions, and does
not have a clear attacker model.
They describe conditional erasure policies where the condition is 
 independent of the data to be erased.
Although this
appears similar to Type~1
policies (restricted to total erasure), it is more accurately
viewed as a form of Type~0 policy in which the
condition defines the point in the trace from which the attacker
begins observation. 

The present paper does not model the behaviour of the user who interacts with an erasing
system. This was studied in \cite{DelTedesco:Sands:Secco09} for one
particular system and attacker model.  We believe that it would be
possible to extend the system model with a user-strategy parameter
(see \cite{Wittbold:Johnson:Information,str1,str1} which consider explicit models of user strategies).
Neither do we consider here the verification or enforcement of erasure
policies; for specific systems and attacker models this has
been studied in a programming language context in
\cite{Hunt:Sands:ESOP08,Chong:PhD,Chong:Myers:End,DelTedesco+:Implementing,Nanevski+:Verification}.


\paragraph{Knowledge based approaches}
%
Our use of knowledge
sets was inspired by Askarov and Sabelfeld's \emph{gradual release} definitions \cite{Askarov:2007}. 
This provides a clear
attacker-oriented perspective on information-flow properties based on
what an attacker can deduce about a secret after making
observations. A number of recent papers have followed this approach to
provide semantics for richer information flow properties, 
e.g. \cite{Banerjee08expressivedeclassification,Broberg:Sands:PLAS09}.
Our use of knowledge sets to build a K-space,
thus generalising the use of equivalence relations/partitions,
is new.
The use of partitions in expressing a variety of information
flow properties was studied in early work by Cohen
\cite{Cohen:InformationB}. The use of equivalence relations and
more generally partial equivalence relations as models for information
and information flow was studied in \cite{Landauer:Redmond:CSFW93} and
resp.~\cite{Sabelfeld:Sands:Per}. 

Recent work \cite{Balliu:2011} uses an epistemic temporal logic
as a specification language for information flow policies.
Formulae are interpreted over trace-based models of programs
in a simple sequential while language (without input actions),
together with an explicit observer
defined via an observation function on traces.
Our work looks very similar in spirit to \cite{Balliu:2011},
though this requires further investigation,
and it appears that our modelling capabilities are comparable.
The use of temporal logic in \cite{Balliu:2011} is attractive,
for example because of the possibility of using off the shelf model-checking tools.
However, our policy language allows
a more intuitive reading and clear representation of the information leakage. 
   
 
Alur \emph{et al} \cite{Alur06preservingsecrecy}, study preservation of secrecy
under refinement. 
The
information flow model of that work bears a number of similarities
with the present work. Differences include a more concrete treatment
of traces, and a more abstract treatment of secrets. As here, equivalence relations are used to
model an attacker's observational power, 
while knowledge models the ability of an attacker to determine the
value of trace predicates.  Their core definition of secrecy coincides
with what we call secrecy of queries
(viz., negation of ``can answer''), although they do not consider
counterparts to our other knowledge-based properties. 

%
%

\paragraph{Abstract Non-Interference}
Abstract Non-Interference \cite{Giacobazzi:Mastroeni:POPL04} has strong similarities
with our use of K-spaces.
In abstract non-interference,
\emph{upper closure operators} (uco's)  are used
to specify non-interference properties.
The similarities with the current work become apparent
when a uco is presented as a \emph{Moore family},
which may be seen as a K-space closed under intersection.

\cite{Giacobazzi:Mastroeni:POPL04} starts by defining the intuitive
notion of \emph{narrow} abstract non-interference (NANI)
parameterized by two upper closure operators $\eta$ (specifying what the attacker can observe of low inputs) and $\rho$ (ditto low outputs).
A weakness of NANI is that it suffers from ``deceptive flows'', whereby a program failing
to satisfy NANI might still be non-interfering.
From our perspective, the deceptive flows problem arises because $\eta$ fails to
distinguish between what an attacker can \emph{observe} of low inputs and what he should
be allowed to \emph{deduce} about them (i.e., everything).
Since we specify the attacker model independently from the flow policy,
the deceptive flows problem does not arise for us.

The deceptive flows problem is addressed in \cite{Giacobazzi:Mastroeni:POPL04}
by defining a more general notion of abstract non-interference (ANI) which introduces
a third uco parameter $\phi$. The definition of ANI adapts that of NANI
by lifting the semantics of a program to an abstract version
in which low inputs are abstracted
by $\eta$ and high inputs by $\phi$. A potential criticism of this approach
is that an intuitive reading is not clear, since it is based on an
abstraction of the original program semantics. On the other hand,
being based on Abstract Interpretation \cite{Cousot:Cousot:Abstract:Interpretation,Cousot81-1},
abstract non-interference has the potential to leverage very well developed
theory and static analysis algorithms for policy checking and enforcement.
It would therefore be useful to explore the connections further
and to attempt an analysis of the ANI definitions
(see also additional variants in \cite{DBLP:conf/aplas/Mastroeni05})
relating them to more intuitive properties based on knowledge sets.
A starting point could be \cite{DBLP:conf/sas/HuntM05}
which provides an alternative characterisation  of NANI using equivalence relations.

\paragraph{Provenance}
A recent abstract model of \emph{information provenance} \cite{Cheney:provenance}
is built on an information-flow foundation and has a number of
similarities with our model, including a focus on an observer model as
an equivalence relation, and a knowledge-based approach described in
terms of queries that an observer can answer.  Provenance is primarily
concerned with a providing \emph{sufficient} information to
answer provenance-related questions. In secrecy and erasure one is
concerned with \emph{not} providing more than a certain amount. 


\section{Conclusions and further work}

We have presented a rich, knowledge-based abstract framework for
erasure policy specification,
taking into account both quantitative and conditional aspects of the problem.
Our model includes an explicit representation of the attacker.
The knowledge-based approach guarantees an intuitive understanding of what it means for an attacker to deduce some information about the secret, and for a policy to provide an upper bound to these deductions.

Our work so far suggests a number of possible extensions. 
At this stage, the most relevant ones on the theoretical side are:
\begin{itemize}

\item Develop a logic defined on traces, both to support policy definition and to give the basis for an enforcement mechanism (as is done in \cite{Balliu:2011}).

\item Model multilevel erasure, based on the fact the attacker might perform observations up-to a certain level in the security lattice. It would be interesting to investigate
different classes of such attackers and to analyse their properties. 

\item Generalise policy specifications to use K-spaces in place of equivalence
relations. This would allow specification of disjunctive policies such as
``reveal the key or the ciphertext, but not both''.
Non-ER policies may also
be more appropriate for protection of facts, rather than queries,
since ER's are effectively closed under complementation and so cannot reveal
a fact without also revealing its negation
(for example, we may be prepared to reveal ``not HIV positive'' to
an insurance company, but not the negation of this fact).

\item Extend the scope of the approach
along the following key \emph{dimensions} (defined in the same spirit as 
\cite{Sabelfeld:Sands:Dimensions}):
\begin{description}

\item[What:] Our model is possibilistic but
it is well known that possibilistic security guarantees can be very weak
when non-determinism is resolved probabilistically
(see the example in Section~5 of \cite{Sabelfeld:Sands:ESOP99}).
A probabilistic approach would be more expressive and provide stronger guarantees.

\item[When:] Our policies support history-based erasure conditions
but many scenarios require reasoning about the future
(``erase this account in 3 weeks'').
This would require a richer semantic setting in which time is modelled more explicitly.

\item[Who:] We do not explicitly model the user's behaviour but it is implicit
in our possibilistic approach that the user behaves non-deterministically and,
in particular, that later inputs are chosen independently of the
erasure subject. Modelling user behaviour explicitly would allow us to relax
this assumption (which is not realistic in all scenarios) and also 
to model
active attackers.
\end{description}
\item 
  Understand the interplay between erasure and cryptographic
  concepts. To make this possible some refinements of the theory are
  needed. Firstly, it would be natural to move to a probabilistic
  system model. Secondly, the present notion of knowledge 
  assumes an attacker with computationally unlimited deductive power;
  instead we would need a notion of feasibly computable knowledge.

\end{itemize}


We have focussed on characterising expressive erasure policies, but not
on their verification for actual systems. 
As a step towards bridging this to more practical
experiments in information erasure, it would be instructive to explore
the connections to the rich policies expressible by the enforcement
mechanism for Python programs we describe in our earlier work
\cite{DelTedesco+:Implementing}.  
\bibliographystyle{splncs03} 
\bibliography{bibliography,literature} 
\normalsize

\appendix 
\section{Proofs}
\begin{lemma}\label{lemma:union-preserves-korder}
Let $I$ be a non-empty index set.
Let $\{ W_i \}_{i \in I}$ be a family of non-empty sets such that $\bigcup_{i \in I} W_i = V$.
Let $\{K_i\}_{i \in I}$ and $\{K'_i\}_{i \in I}$ be families
of K-spaces, with each $K_i, K'_i$ covering $W_i$.
Then, for $o \in \{ \oname{L}, \oname{U}, \oname{EM}, \oname{CA}, \oname{WA} \}$:
\[
	(\forall i \in I . K_i \mathrel{{\korder}_o} K'_i)
	=> \bigcup_{i \in I}K_i \mathrel{{\korder}_o} \bigcup_{i \in I}K'_i
\]

\begin{proof}
We show the two interesting cases, $\oname{CA}$ and $\oname{WA}$.
\begin{itemize}
\item case $\oname{CA}$. Assume $\forall i \in I . K_i \mathrel{{\korder}_\oname{CA}} K'_i$ and consider a query $Q \subseteq V$ such that $\bigcup_{i \in I}K'_i$ can answer $Q$. By definition this implies there exists a $j \in I$ such that $\exists X' \in K_j'$ and either $X' \subseteq Q$ or $X' \subseteq V \setminus Q$.
\begin{itemize}
\item Suppose $X' \subseteq Q$, then $Q'=W_j \cap Q$ is a query $K_j'$ can answer via $X'$. Since $K_j \mathrel{{\korder}_\oname{CA}} K_j'$, $K_j$ can answer $Q'$ as well, therefore there must be a $X \in K_j$ such that either $X \subseteq Q'$ or $X \subseteq  W_j \setminus Q'$. If  $X \subseteq Q'$ then $\bigcup_{i \in I}K_i$ can answer $Q$ via $X$ in $K_j$. Otherwise $X  \subseteq W_j \setminus Q'$, but this means $X \subseteq V \setminus Q$ therefore $\bigcup_{i \in I}K_i$ can answer $Q$ via $X$ in $K_j$.
\item Suppose $X' \subseteq V \setminus Q$, then  $Q'=W_j \setminus Q$ is a query $K_j'$ can answer via $X'$. For the same reason we explained previously, there must be a $X \in K_j$ such that either $X\subseteq Q'$ or $X \subseteq W_j \setminus Q'$. If $X \subseteq Q'$ then $X \subseteq V \setminus Q$ and $\bigcup_{i \in I}K_i$ can answer $Q$ via $X$ in $K_j$. Otherwise $X \subseteq W_j \setminus Q'$, but this means $X \subseteq Q$ therefore $\bigcup_{i \in I}K_i$ can answer $Q$ via $X$ in $K_j$.
\end{itemize}
\item case $\oname{WA}$. Assume $\forall i \in I . K_i \mathrel{{\korder}_\oname{WA}} K'_i$ and consider a query $Q \subseteq V$ such that $\bigcup_{i \in I}K'_i$ will answer $Q$. By definition this implies that $\forall X' \in K_j'$, either $X' \subseteq Q$ or $X' \subseteq V \setminus Q$ for all $K_j'$ in $\{K'_i\}_{i \in I}$. Let us consider one $K_j'$ of the family and define $Q_j'=Q \cap W_j$. Then we have $\forall X' \in K_j'$, either $X' \subseteq Q_j'$ or $X' \subseteq W_j \setminus Q_j'$, therefore $Q_j'$ is a query $K_j'$ will answer. Since $K_j \mathrel{{\korder}_\oname{WA}} K_j'$, $K_j$ will answer $Q_j'$ as well, therefore $X \subseteq Q_j'$ or $X \subseteq W_j \setminus Q_j'$ must hold for all $X \in K_j$. But this implies $X \subseteq Q$ or $X \subseteq V \setminus Q$ as well, and the statement holds for all $K_j$ in  $\{K_i\}_{i \in I}$, therefore $\{K_i\}_{i \in I}$ will answer $Q$.
\end{itemize}\qed
\end{proof}
\end{lemma}

\begin{lemma}\label{lemma:conjunction-vs-union}
Let $\{P_i\}_{i \in I}$ be a non-empty family of relations in $\ER(V)$ for some set $V$,
and let $R = \bigcap_{i \in I}P_i$.
Then $\class{R} \korder[EM] \bigcup_{i \in I}\class{P_i}$.
\begin{proof}
Every element of $\class{R}$ is of the form $\class{v}_R$,
every element of $\bigcup_{i \in I} \class{P_i}$ is of the form $\class{v}_{P_i}$ for some
$i \in I$,
and every choice of $v$ and $i$ generates such elements.
It thus suffices to show that
$\class{v}_R \subseteq \class{v}_{P_i}$ for all choices of $v$ and $i$. This follows since
$R$ is a finer equivalence relation than each $P_i$.\qed
\end{proof}
\end{lemma}

\begin{lemma}\label{lemma:emorder-with-union}
Let $P \in PT(T_V)$, $S \in Sys(V)$ and $A \in ER(T)$. Then $\bigcup_{X \in \class{P}}$ $\kspace{S\cap X}{A} \korder[EM] \kspace{S}{A}$.
\begin{proof}
We show the lower and the upper ordering separately.

\begin{enumerate}
\item $\forall X \in \class{P} . \kspace{S\cap X}{A} \korder[L] \kspace{S}{A}$:

Let $Y \in \kspace{S\cap X}{A}$. Then, for some $O \in \class{A}$,
$Y = \subject(O \cap S \cap X)$ and $O \cap S \cap X$ is non-empty.
Let $Y' = \subject(O \cap S)$. Then $O \cap S$ is non-empty (since $O \cap S \cap X$ is non-empty), hence $Y \subseteq Y' \in \kspace{S}{A}$.

\item $\forall Y \in \kspace{S}{A}.
		\exists X \in \class{P} .
		\exists Y' \in \kspace{S\cap X}{A} . Y' \subseteq Y$:

Let $Y \in \kspace{S}{A}$. Then, for some $O \in \class{A}$,
$Y = \subject(O \cap S)$ and $O \cap S$ is non-empty.

Suppose, towards a contradiction, that
$O \cap S \cap X = \emptyset$ for all $X \in \class{P}$, hence
$O \cap S \cap \bigcup_{X \in \class{P}} X = \emptyset$; but
$\class{P}$ partitions $T$, so $\bigcup_{X \in \class{P}} X = T \supseteq O \cap S$,
hence $O \cap S = \emptyset$, a contradiction.

So let $X \in \class{P}$ with $O \cap S \cap X$ non-empty, and let
$Y' = \subject(O \cap S \cap X)$. Then $Y' \subseteq Y$ and $Y' \in \kspace{S\cap X}{A}$.
\end{enumerate}

From 1 it follows that
$\bigcup_{X \in \class{P}} \kspace{S\cap X}{A} \korder[L] \kspace{S}{A}$
and from 2 it follows that
$\bigcup_{X \in \class{P}} \kspace{S\cap X}{A} \korder[U] \kspace{S}{A}$,
thus:
\[
	\bigcup_{X \in \class{P}} \kspace{S\cap X}{A} \korder[EM] \kspace{S}{A}
\]
\end{proof}
\end{lemma}

\subsection{Proof of Therorem \ref{theorem:explicitness-type1}}
By Lemma~\ref{lemma:emorder-with-union}, $\bigcup_{X \in \class{P}} \kspace{S\cap X}{A} \korder[EM] \kspace{S}{A}$.

By assumption of policy satisfaction, $\class{f\;X} \mathrel{{\korder}_o} \kspace{S\cap X}{A}$ for all $X \in \class{P}$,with each $\kspace{S\cap X}{A}$ covering $V$.

So, by Lemma~\ref{lemma:union-preserves-korder}:
\[ 
	\bigcup_{X \in \class{P}} \class{f\;X}
	\mathrel{{\korder}_o} \bigcup_{X \in \class{P}} \kspace{S\cap X}{A}
\]

It then suffices to show that
$\class{\bigcap_{X \in \class{P}}(f\;X)} \korder[EM] \bigcup_{X \in \class{P}} \class{f\;X}$.
This is immediate by Lemma~\ref{lemma:conjunction-vs-union}.\qed

\begin{lemma}\label{lemma:properties-dagger}:
Let $\{ R_W \}_{W \in Q}$ be a partition-indexed family of equivalence relations
such that $R_W \in \ER(W)$ for each $W \in Q$. Then:
\begin{enumerate}

\item $\bigwedge_{W \in Q}R^{\dagger}_{W} = \bigcup_{W \in Q}R_{W}$

\item $\bigwedge_{W \in Q}R^{\dagger}_{W} \subseteq {\erpart{Q}}$

\end{enumerate}
\begin{proof}
\begin{enumerate}
\item Recall  $R^{\dagger}_W = R_W \cup \All_{V\setminus W}$.
Then, for all $W$ in the partition $Q$, $\forall (x,y) \in R^{\dagger}_{W}$ either $x \in W \wedge y \in W$ or $x \not \in W \wedge y \not \in W$. In fact, suppose $x \in W$ but $y \not \in W$, then $(x,y) \not \in R_W$ because $y \not \in W$ and $(x,y)\not \in \All_{V\setminus W}$ because $x \not  \in V\setminus W$, a contradiction.

We now show $\bigwedge_{W \in Q}R^{\dagger}_{W} \subseteq \bigcup_{W \in Q}R_{W}$. Consider  $(x,y) \in \bigwedge_{W \in Q}R^{\dagger}_{W}$. Then $\exists W\in Q. x \in W \wedge y\in W$ because of the previous result, therefore $(x,y) \in R_{W}\subseteq \bigcup_{W \in Q}R_{W}$.

We now show $\bigwedge_{W \in Q}R^{\dagger}_{W} \supseteq \bigcup_{W \in Q}R_{W}$. Consider $(x,y) \in \bigcup_{W \in Q}R_{W}$. Since $\forall W,W' \in Q. W\cap W'=\emptyset$, $\exists W \in Q. (x,y) \in R_{W}$, therefore $(x,y) \in R^{\dagger}_{W}$. For all others $W' \in Q. W'\not = W$ we have $x \not \in W' \wedge y \not \in W'$, therefore $(x,y) \in R^\dagger_{W'}$. So we can conclude $(x,y) \in \bigwedge_{W \in Q}R^{\dagger}_{W}$

\item Consider $(x,y)\in \bigwedge_{W \in Q}R^{\dagger}_{W}$. Since $\bigwedge_{W \in Q}R^{\dagger}_{W} = \bigcup_{W \in Q}R_{W}$ and since $\forall W,W' \in Q. W\cap W'=\emptyset$, $\exists ! W \in Q . (x,y) \in R_{W}$. But $R_{W} \subseteq W \times W$ and $W \times W \subseteq \erpart{Q}$ by definition, therefore $(x,y)\in \erpart{Q}$.  
\end{enumerate}\qed
\end{proof}
\end{lemma}

\subsection{Proof of Therorem \ref{theorem:explicitness-type2}}
Recall the definition of $T_W=\{t \in T|\Phi(t)\in W\}$. 

Let $P_Q$ be a partition of $T_V$ defined as $P_Q=\bigcup_{W \in \class{Q}}T_W$. 

By Lemma~\ref{lemma:emorder-with-union} we have $\bigcup_{T_W \in [P_Q]} \kspace{S\cap T_W}{A} \korder[EM] \kspace{S}{A}$.

We then have $\class{R_{(g\;W)}} \mathrel{{\korder}_o} \kspace{S \cap T_W }{A}$ for all $T_W  \in [P_Q]$ by assumption of policy satisfaction and Theorem \ref{theorem:explicitness-type1}
applied to all subsystems $S \cap T_W$.

By Lemma \ref{lemma:properties-dagger} we have $\erpart{Q} \wedge \bigwedge_{W \in Q}R^{\dagger}_{(g\;W)} = \bigwedge_{W \in Q}R^{\dagger}_{(g\;W)} = \bigcup_{W \in Q}R_{(g\;W)}$.

To conclude the proof we only need $\bigcup_{W \in Q}\class{R_{(g\;W)}} \mathrel{{\korder}_o} \bigcup_{T_W \in P_Q}\kspace{S \cap T_W}{A}$, which holds by  Lemma~\ref{lemma:union-preserves-korder} .\qed

\end{document}